\documentclass[reqno, 10pt]{amsart}

\usepackage{graphicx, verbatim, amsmath, amssymb, mathrsfs, cite, xspace, scrextend}
\usepackage{color}
\usepackage{fullpage}
\usepackage[latin1]{inputenc}%
\usepackage{amsmath}
\newtheorem{defn}[subsection]{Definition}

\newcommand{\F}{\mathbb{F}}
\begin{document}

\title{Ironwood Meta Key Agreement and Authentication Protocol}

\author{Iris Anshel}
\address{SecureRF Corporation, 100 Beard Sawmill Rd \#350, Shelton, CT 06484}
  \email{ianshel@securerf.com}
  
  \author{Derek Atkins}
\address{SecureRF Corporation, 100 Beard Sawmill Rd \#350, Shelton, CT 06484}
  \email{datkins@securerf.com}

\author{Dorian Goldfeld}
\address{SecureRF Corporation, 100 Beard Sawmill Rd \#350, Shelton, CT 06484 }
\email{dgoldfeld@securerf.com}

\author{Paul E. Gunnells}
\address{SecureRF Corporation, 100 Beard Sawmill Rd \#350, Shelton, CT 06484}
\email{pgunnells@securerf.com}
%\author{\IEEEauthorblockN{Iris Anshel\IEEEauthorrefmark{1},
%Derek Atkins\IEEEauthorrefmark{2}, Dorian Goldfeld\IEEEauthorrefmark{3} and
%Paul E. Gunnells\IEEEauthorrefmark{4} \\}
%\IEEEauthorblockA{SecureRF Corporation,
%100 Beard Sawmill Rd \#350, Shelton, CT 06484 \\
%Email: \IEEEauthorrefmark{1}ianshel@securerf.com,
%\IEEEauthorrefmark{2}datkins@securerf.com,
%\IEEEauthorrefmark{3}dgoldfeld@securerf.com,
%\IEEEauthorrefmark{4}pgunnells@securerf.com}}

\maketitle

\begin{abstract}
Number theoretic public-key solutions, currently used in many
applications worldwide, will be subject to various quantum attacks,
making them less attractive for longer-term use.  Certain group
theoretic constructs are now showing promise in providing
quantum-resistant cryptographic primitives, and may provide suitable
alternatives for those looking to address known quantum attacks.  In
this paper, we introduce a new protocol called a \emph{Meta Key
Agreement and Authentication Protocol} (MKAAP) that has some
characteristics of a public-key solution and some of a shared-key
solution.  Specifically it has the deployment benefits of a public-key
system, allowing two entities that have never met before to
authenticate without requiring real-time access to a third-party, but
does require secure provisioning of key material from a trusted key
distribution system (similar to a symmetric system) prior to
deployment.  We then describe a specific MKAAP instance, the Ironwood
MKAAP, discuss its security, and show how it resists certain quantum
attacks such as Shor's algorithm or Grover's quantum search algorithm.
We also show Ironwood implemented on several ``internet of things''
(IoT devices), measure its performance, and show how it performs
significantly better than ECC using fewer device resources.

\end{abstract}

\section{Introduction}

Group theoretic cryptography is a relatively new discipline that seeks
to bring the core algorithmically difficult problems in combinatorial
group theory into the cryptographic landscape. Overviews can be found
in the two recent monographs \cite{GMS}, \cite{MU2}.  Among the first
generation of group theoretic based key agreement protocols to be
introduced, including \cite{AAG} and \cite{KLCHKP}, were those based
of the conjugacy search problem. The attacks on the conjugacy search
problem, such as those appearing in \cite{GKTTV}, \cite{G}, \cite{HS}
suggest that these types of schemes may not be practical over braid
groups in low-resource environments. To overcome these concerns, we
introduce the notion of a \emph{Meta Key Agreement and Authentication
Protocol} (MKAAP) (\S \ref{MKAAP}).  This protocol has many of the properties
and advantages of a public-key method and requires very limited
distribution of certain private keys.

In this paper we present an MKAAP based on a conjectured
quantum-resistant one-way function based in braid group theory.  To
date, this MKAAP is immune to all known attacks introduced in group
theoretic cryptography and delivers linear time performance on
low-footprint processors.  As the IoT becomes ubiquitous the need to
secure low-footprint processors can be achieved using group theoretic
cryptography via the MKAAP introduced below.
 
\subsection*{Previous Work}

In 2006 \cite{aagl} in joint work with M.~Anshel and S.~Lemieux two of
us (IA and DG) introduced a key agreement protocol based in group
theory (specifically the braid group) that has withstood several
attacks over the past decade.  First Myasnikov--Ushakov \cite{MU}
determined that if braids are too short then one can find the
conjugating factor and use that to break the system.  However it was
pointed out by one of us (PG)\cite{Gu} that in practice the braids are
long enough that this attack can never succeed: the method in
\cite{MU} is analogous to using Fermat's technique to factor short RSA
keys, which becomes impractical at secure sizes.  Second,
Kalka--Teicher--Tsaban \cite{KTT} described a linear algebra attack
(KTT) that would allow an attacker to determine part of the private
key data.  However, two of us (DG and PG)\cite{GG} showed that this
attack succeeds only on a class of weak keys, and that choosing the
private key data more carefully defeats this attack.  Subsequent to
the KTT attack, Ben-Zvi--Blackburn--Tsaban \cite{BBT}, using all of
the available public information of the protocol, were able to
reconstruct the shared secret, after a large precomputation and
several hours of runtime.  We later showed \cite{aagg} that the work
necessary to carry out the attack increases as the size of the
permutation order grows as well as the size of the braid group.
 
We remark that the current review of WalnutDSA \cite{WalnutDSA}, a
group theoretic based digital signature, does not apply to the
Ironwood protocol.  In particular the (exponential) attack on
reversing E-multiplication requires data not available to an attacker,
and hence the underlying hard problems considered in these approaches
do not impact the Ironwood security (see \S VI).

\subsection*{Our Contribution}

This paper introduces the \emph{Ironwood MKAAP}, whose security is
based on hard problems in group theory.  Ironwood leverages the
conjectured one-way function, E-Multiplication, but creates a
different construction that removes some of the public information
required to mount any of the previous attacks.  In addition to being
immune from previous attacks, it can be argued that Ironwood is
resistant to the current generation of quantum attacks.  Specifically,
Shor's quantum algorithm\cite{shor} -- which has been shown to break
RSA, ECC, and several other public-key crypto systems -- does not seem
applicable for attacking Ironwood because the underlying group
theoretic foundation of Ironwood is an infinite non-commutative group.
Further, Grover's quantum search algorithm\cite{grover} is not as
impactful on Ironwood as it is on many protocols since the running
time of Ironwood is linear in the key length and security strength,
and hence the need to double the security level amounts to doubling
the execution time. This stands in stark contrast to the rapid
increase in execution time for standard protocols when the lengths of
the private keys are doubled.

This paper begins with a review of the braid group, the colored Burau
representation, and E-Multiplication. With these tools in place, we
introduce the concept of a meta key agreement and authentication
protocol (MKAAP), and present Ironwood. A discussion of security and
our implementation experience follows.

\section{Colored Burau representation of the braid group}

Let $B_N$ denote the braid group on $N$ strands, with Artin
presentation

$$B_N = \Bigl\langle b_1,b_2, \ldots, b_{N-1} \; \Bigm |  b_i b_j b_i =  b_j b_i b_j  \quad \text{for} \; |i-j|=1,\quad  
b_i b_j =  b_j b_i    \quad \text{for} \; |i-j| \ge 2 \Bigr\rangle.$$

Let $S_{N}$ be the permutation group on $N$ letters.  Every element
$\beta \in B_{N}$ determines a permutation $\sigma_\beta \in S_{N}$ as
follows.  For $1\le i < N$, let $\sigma_i \in S_N$ be simple
transposition that exchanges $i$ and $i+1$ and leaves the remaining
elements $\{1,\ldots,i-1,i+2,\ldots,N\}$ fixed. We write $\sigma_{b_i}
= \sigma_i$. Then if $\beta = b_{ i_1}^{\epsilon_1} \; b_{i_2
}^{\epsilon_2}\; \cdots\; b_{i_k }^{\epsilon_k },$ (with $\epsilon_i =
\pm 1$), we have $\sigma_\beta = \sigma_{i_1}\dotsb \sigma_{i_k}$.

The colored Burau representation of the braid group was introduced by
Morton in \cite{M} in 1998, but we shall make use of a variation of
Morton's original representation.  Associate to each Artin generator
$b_i$, with $1\le i<N$, a colored Burau matrix $CB(b_i)$ where
\begin{equation}\label{eq:cbmatrix}
\begin{aligned}
& CB (b_{1}) = \left(\begin{array}{ccccc}
-t_1& 1&&&\\
&\ddots&&&\\
&&1&&\\
&&&\ddots&\\
&&&&1
\end{array} \right), \\
& CB (b_{i}) = \left(\begin{array}{ccccc}
1&&&&\\
&\ddots&&&\\
&t_{i}&-t_{i}&1&\\
&&&\ddots&\\
&&&&1
\end{array} \right) \; \\ & \big(\text{for} \; 1<i<N\big).\end{aligned}\end{equation}

We similarly define $CB (b_{i}^{-1})$ by modifying \eqref{eq:cbmatrix} slightly:

\begin{equation}\label{eq:cbmatrix2}
\begin{aligned}
& CB(b_{1}^{-1}) = \left(\begin{array}{ccccc}
 -\frac{1}{t_2}& \frac{1}{t_2}&&&\\
&\ddots&&&\\
&&1&&\\
&&&\ddots&\\
&&&&1
\end{array} \right), \\
& CB (b_{i}^{-1}) = \left(\begin{array}{ccccc}
1&&&&\\
&\ddots&&&\\
&1&-\frac{1}{t_{i+1}}&\frac{1}{t_{i+1}}&\\
&&&\ddots&\\
&&&&1
\end{array} \right) \; \\ & \big(\text{for} \; 1<i<N\big).\end{aligned}\end{equation}

Thus each braid generator $b_i$ (respectively, inverse generator
$b_{i}^{-1}$) determines a colored Burau/permutation pair
$(CB(b_i),\sigma_i )$ (resp., $(CB (b_{i}^{-1}), \sigma_{i})$).
We now wish to define a multiplication
of colored Burau pairs such that the natural mapping from the braid
group to the group of matrices with entries in the ring of Laurent
polynomials in the $t_{i}$ is a homomorphism.

Given a Laurent polynomial $$f(t_1,\ldots,t_N) \;\in\; \Bbb Z[t_1^{\pm 1}, t_2^{\pm 1}, \ldots, t_N^{\pm 1}],$$ a
permutation in $\sigma \in S_N$ can act (on the left) by permuting the
indices of the variables.  We denote this action by $f\mapsto \!\!\!\phantom{a}^{\sigma}f$:
$$
^{\sigma} f(t_1,t_2,\ldots,t_N )\; := \;f(t_{\sigma(1)} ,t_{\sigma(2)} ,\ldots,t_{\sigma(N)} ).
$$
Let $\mathcal{M}$ be the $N \times N$ matrices over $\Bbb Z[t_1^{\pm
1}, t_2^{\pm 1}, \ldots, t_N^{\pm 1}]$.  We extend this action to
${\mathcal M}$ by acting on each entry in a matrix, and use the same
notation for the action.  The general definition for multiplying two
colored Burau pairs is now defined as follows from the definition of
the semidirect product ${\mathcal M} \rtimes S_N $. Given
$b_i^{\pm},b_j^{\pm}$, the colored Burau/permutation pair associated
with the product $b_i^{\pm}\cdot b_j^{\pm}$ is
$$(CB(b_i^{\pm}),\sigma_i ) \circ (CB(b_j^{\pm}), \; \sigma_j ) = 
 \Big(CB(b_i^{\pm} ) \cdot ( ^{\sigma_i} CB(b_j^{\pm} )), \;
\sigma_i\cdot \sigma_j \Big).$$
Given any braid 
$$\beta = b_{i_1}^{\epsilon_1} b_{i_2}^{\epsilon_2} \cdots b_{i_k}^{\epsilon_k},$$ 
with $\epsilon_i = \pm 1$ for $1\le i \le k$,  the colored Burau pair 
 $(CB(\beta),\sigma_\beta)$ is given by
 
$$ (CB(\beta),\sigma_\beta) = \left(CB(b_{i_1}^{\epsilon_1} ) \cdot
^{\sigma_{i_1}} CB(b_{i_2} ^{\epsilon_2}) \cdot
^{\sigma_{i_1}\sigma_{i_2}} CB(b_{i_3}^{\epsilon_3}) )\;\; \cdots \;\;
^{\sigma_{i_1}\sigma_{i_2}\cdots \sigma_{i_{k-1}}}
CB(b_{i_k}^{\epsilon_k}), \;\;\sigma_{i_1}\sigma_{i_2}\cdots
\sigma_{i_{k}} \right).
$$
The colored Burau representation is then defined by
$$\Pi_{CB}(\beta) := (CB(\beta),\sigma_\beta).$$ One checks that
$\Pi_{CB}$ satisfies the braid relations and, hence, defines a
representation of $B_N.$

\section{E-Multiplication}

  E-Multiplication was first introduced in \cite{aagl} as a core building block for a range of cryptographic constructions.  We recall its definition here.
     
    Let $\F_{q}$ denote the finite field of $q$ elements.  A set of $T$-values is defined to be a collection of non-zero field elements:
$$\{ \tau_1,\tau_2,\ldots,\tau_N \}\subset \F_q^{\times }.$$ Given a set of
$T$-values, we can evaluate any Laurent
polynomial $f(t_1,t_2,\dots,t_N) $ to obtain an element of $\F_{q}$:
$$f(t_1,t_2,\ldots,t_N ) \downarrow_{t\text{\rm -values}} \; :=   f(\tau_1,\tau_2,\ldots,\tau_N).$$
We extend this notation to matrices over Laurent polynomials in the
obvious way.  

With all these components in place, we can now define E-Multiplication. 
By definition, E-Multiplication is an operation that takes as input two ordered pairs, 
$$(M,\sigma_0 ), \quad (CB(\beta),\sigma_\beta ),$$ where $\beta \in
B_{N}$ and $\sigma_{\beta} \in S_{N}$ as before, and where $M\in GL(N,
\F_q)$, and $\sigma_0 \in
S_N$. We denote E-Multiplication with a star: $\star$. The result of E-Multiplication, denoted
$$(M',\sigma' ) = (M,\sigma_0 )\star (CB(\beta),\sigma_\beta ),$$ will
be another ordered pair $(M',\sigma' )\in GL(N, \F_q)\times S_{N}$.  

We define E-Multiplication inductively.  When the braid $\beta =
b_i^{\pm}$ is a single generator or its inverse, we put
$$(M,\sigma_0 )\star \big(CB(b_i^{\pm}\big), \;\sigma_{b_i^{\pm} } ) = $$
$$\;\;\;\;\;\Big(M\cdot\, ^{\sigma_0} \big(CB(b_i^{\pm} \big))
\downarrow_{t\text{\rm -values}}, \;\; \sigma_0\cdot \sigma_{b_i^{\pm}
}\Big).$$ In the general case, when $\beta = b_{i_1}^{\epsilon_1}
b_{i_2}^{\epsilon_2} \cdots b_{i_k}^{\epsilon_k},$ we put
\begin{equation}\label{eq:emultstar}(M,\sigma_0 )\star (CB(\beta),\sigma_\beta) = 
 (M,\sigma_0 )\star (CB(b_{i_{1}}^{\epsilon_1}),\sigma_{b_{i_{1}} })\star (CB(b_{i_{2}}^{\epsilon_2}),\sigma_{b_{i_{2}} })\star  
 \cdots \star (CB(b_{i_{k}}^{\epsilon_k}),\sigma_{b_{i_{k}} }),
\end{equation}
where we interpret the right of \eqref{eq:emultstar} by associating
left-to-right.  One can check that this is independent of the
expression of $\beta$ in the Artin generators.

\section{Meta  Key Agreement and Authentication Protocol (MKAAP) \label{MKAAP}}
 
We now introduce the notion of a meta key agreement and authentication
protocol which is not a true public-key crypto system but has many of
the features of a public-key cryptosystem. Specifically, while it does
initially require secure provisioning of each device by a Trusted
Third Party (TTP), after this devices can authenticate to each other
offline without further support.  Here by {\em device} we mean a
Probabilistic Polynomial-Time Turing Machine (PPTM) that can execute a
cryptographic protocol and be capable of transmitting and receiving
messages.
    
\vskip 10pt\noindent {\bf Definition (MKAAP)} Assume there is a
network consisting of a Home Device (HD) and a set of other devices
$D_i$, $i=1, 2, 3, \ldots$ that communicate with the HD over an open
channel.\footnote{While the open channel may be a unicast or multicast
medium, authentication between HD and $D_i$ is 1:1.} Assume that
there is a TTP which has distributed secret information to the HD and
the other devices. An MKAAP is an algorithm with the following
properties:

  {\it   \vskip 6pt   $\bullet$ The  MKAAP  allows the HD to authenticate (and/or be authenticated by) and obtain a shared secret with any $D_i$ over an open channel.
      \vskip 6pt $\bullet$ It is infeasible for an attacker, eavesdropping on the open communication channel between the HD and a device $D_i$, to obtain the shared secret assuming the attacker does not know the secret information distributed by the TTP.
    \vskip 6pt $\bullet$ The private keys of the $D_i$ are provided by the TTP, fixed, and are not known to the HD.  The TTP may update the keys over time.
    \vskip 6pt
    $\bullet$ The private key of the HD may be ephemeral and is not known to any of the $D_i$'s, or it may be provided by the TTP.
    
   \vskip 6pt
   $\bullet$ If an attacker can break into one of the devices $D_i$ and obtain its private key, then only the security of that particular device is breached; all other devices remain secure.
   \vskip 6pt
   $\bullet$ An attacker is assumed to be a Probabilistic Polynomial-Time Turing Machine (PPTM) and/or a machine capable of running a quantum method like Shor\cite{shor} or Grover\cite{grover}, capable of passive eavesdropping on all communications between the HD and $D_i$, and can actively attempt to impersonate an HD or $D_i$, using the other side as an oracle.
    }

\vskip10pt

One may question what the benefit of an MKAAP system is over a pure
symmetric solution.  The MKAAP requires the endpoints to be
provisioned similarly to a symmetric solution.  However, in a
symmetric solution, the HD must be provisioned with the keys for
\emph{every} $D_i$, or at least must have real-time access to the TTP to
obtain such keys.  In MKAAP, however, the HD can be provisioned
\emph{before} keys for the $D_i$ are even generated, and the HD has no
need to contact the TTP.  This allows for additional devices to be
created and provisioned after the HD is already deployed without
requiring any changes to the HD.  This would not be possible with a
symmetric solution.

\section{Ironwood MKAAP}

  We now describe the Ironwood MKAAP. It may be assumed that the following information is publicly known.
  \vskip 10pt
  \noindent
   {\bf Public Information:}

  \vskip 10pt
  $\bullet$ The braid group $B_N$ for a fixed even integer $N \ge 10$.
  
  $\bullet$ A finite field $\F_q$ of $q \ge 7$
  elements.\footnote{The parameters $N$ and $q$ are chosen to meet the desired security level given the best-known attack (which is currently brute force).  See Section \ref{s:implementation}.}
 
 $\bullet$ A non-singular matrix $m_0 \in GL(N, \F_q).$
 
 $\bullet$ The operation of E-multiplication based on $B_N$ and $\F_q$.

\vskip 10pt 
 Next, we discuss the initial distribution of secret information by the TTP.
 
 \vskip 10pt\noindent
 {\bf TTP Data Generation and Distribution:}

\begin{figure}[h!]
  \centering
  \includegraphics[width=0.9\linewidth]{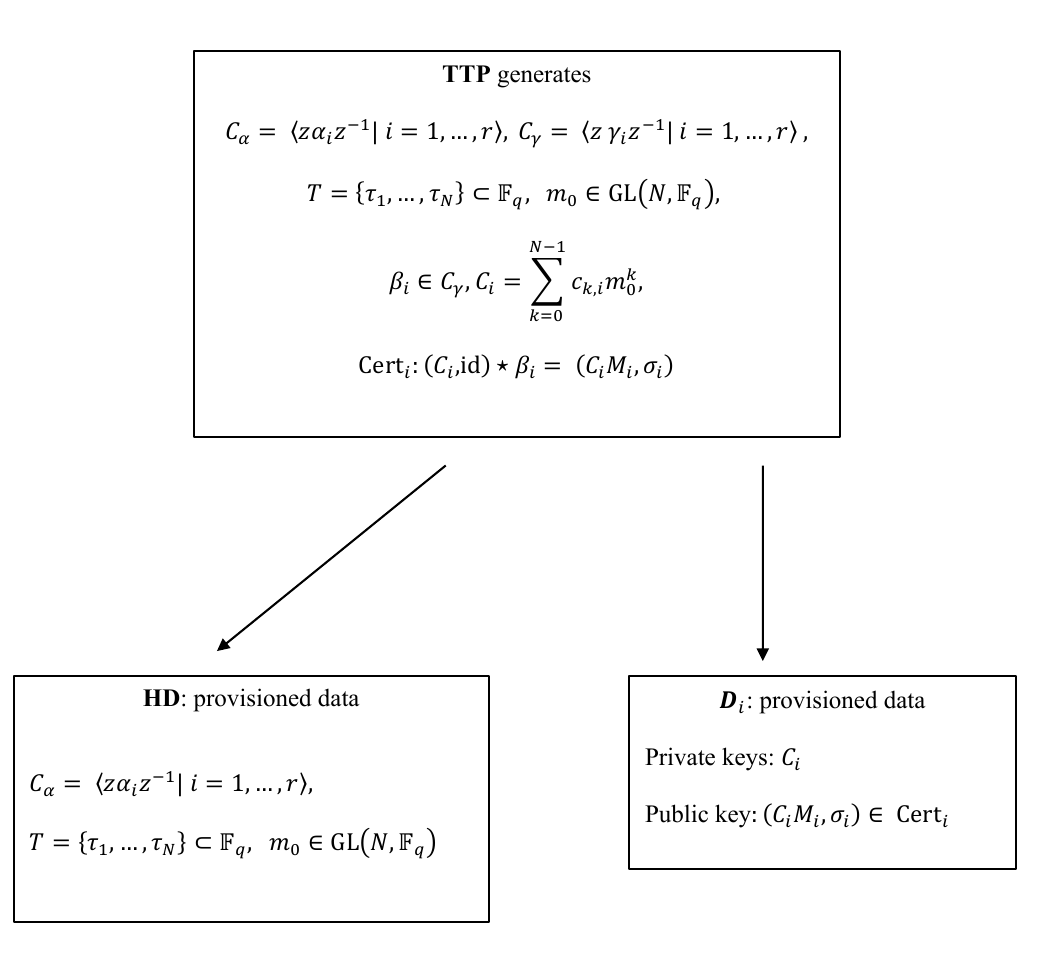}
  \caption{Ironwood Data Flows}
  \label{f:data_flows}
\end{figure}

\vskip 10pt
The TTP begins by generating two sets of braid elements
$\{\alpha_1, \ldots,\alpha_r\}$, $\{\gamma_1, \ldots,\gamma_r\}$ such
that $\alpha_i\gamma_j = \gamma_j\alpha_i.$ By conjugating these sets
of elements by a fixed braid element $z$, the TTP creates two sets of
commuting conjugates $\mathcal{C}_{\alpha}, \mathcal{C}_{\gamma}
\subset B_{N}$:\footnote{The number $r$ of braids in each set is
chosen for a time/space/keysize tradeoff, but must satisfy $r \ge 2$.
A larger $r$ requires more time to generate the conjugates and more
space to store them, but on the other hand requires shorter random
words in them to generate keys.}
\begin{align*}
\mathcal C_\alpha &= \{z\alpha_1z^{-1}, \; z\alpha_2 z^{-1}, \;\; \ldots, \;\; z\alpha_r z^{-1}\},\\
\mathcal C_\gamma &= \{z\gamma_1z^{-1}, \; z\gamma_2 z^{-1}, \;\; \ldots, \;\; z\gamma_r z^{-1}\}.
\end{align*}
In addition to the sets $\mathcal{C}_{\alpha}, \mathcal{C}_{\gamma}$
conjugates above, the TTP produces a fixed set of $T$-values: 
\[
T = \{\tau_1, \tau_2, \ldots,\tau_N\} \subset \F_q^{\times }, \qquad
(\tau_i \ne 1).
\] 
The TTP writes the first set of conjugates $\mathcal C_\alpha$ and the
set of $T$-values into the memory of the HD.  After being provisioned
with $\mathcal C_\alpha$ and $T$ the HD functions independently of the
TTP, and is capable of producing both its own private and public keys.
No further interaction is necessary between the TTP and the HD.

 \vskip 3pt

Next the TTP creates braid words $\beta_i \in B_N$ (for $i
=1,2,\ldots$), which are random products of conjugates from the second
set $\mathcal C_\gamma$ chosen according to the uniform distribution.
The TTP then creates colored Burau pairs $(\beta_i, \sigma_i)$, where
$\sigma_i$ is the permutation associated to $\beta_i$. For each such
$(\beta_i, \sigma_i)$, the TTP chooses a random non-singular matrix
\[
C_i = \sum_{k=0}^{N-1} c_{k,i} m_0^k,
\qquad\qquad \big(\text{with} \; c_{k,i}\in \F_q\big),
\]
where the $c_{k,i}\in \F_q$ are chosen according to the uniform
distribution.  The TTP then uses the $T$-values to perform the
E-multiplication
\begin{equation}\label{eq:emult}
\text{Pub}_i := (C_i, \text{Id}) \star (\beta_i,
\sigma_i) = (C_i M_i, \sigma_i).
\end{equation}
In \eqref{eq:emult}, $\text{Id}$ denotes the identity permutation and
$M_i \in GL(N, \F_q)$. We remark that the one-way nature of
E-multiplication makes it impossible for the TTP to choose the public
key for $D_i$ prior to specifying the private key for $D_i$. Thus the
protocol is entirely distinct from any identity-based protocol.
Finally, the TTP creates a certificate $\text{Cert}_i$ that contains a
digitally signed copy of $\text{Pub}_i$ and writes $\text{Cert}_i$ and
$C_i$ into the memory of $D_i$, the $i^{th}$ device in the network. By
using a digital signature to sign the public key of $D_i$ and to
generate $\text{Cert}_i$, we eliminate concerns about the
man-in-the-middle attack.

   \vskip10pt

After all the data is created, the TTP must securely provision the
Home Device and Other Devices with their respective data.  Figure
\ref{f:data_flows} summarizes the data created by the TTP, and how it
is distributed to the devices.
 
 \vskip 10pt

 Once the TTP distribution is completed authentication and key
agreement between the HD and the other devices in the network may
begin. A key assumption is that there is only one HD and that the
secret information on the HD is secure and cannot be obtained by any
adversary.  The protocol proceeds as follows:
 
 \vskip 10pt\noindent
  {\bf Ironwood Authentication and Key Agreement Protocol}

  \begin{figure}[h!]
  \centering
  \includegraphics[width=0.9\linewidth]{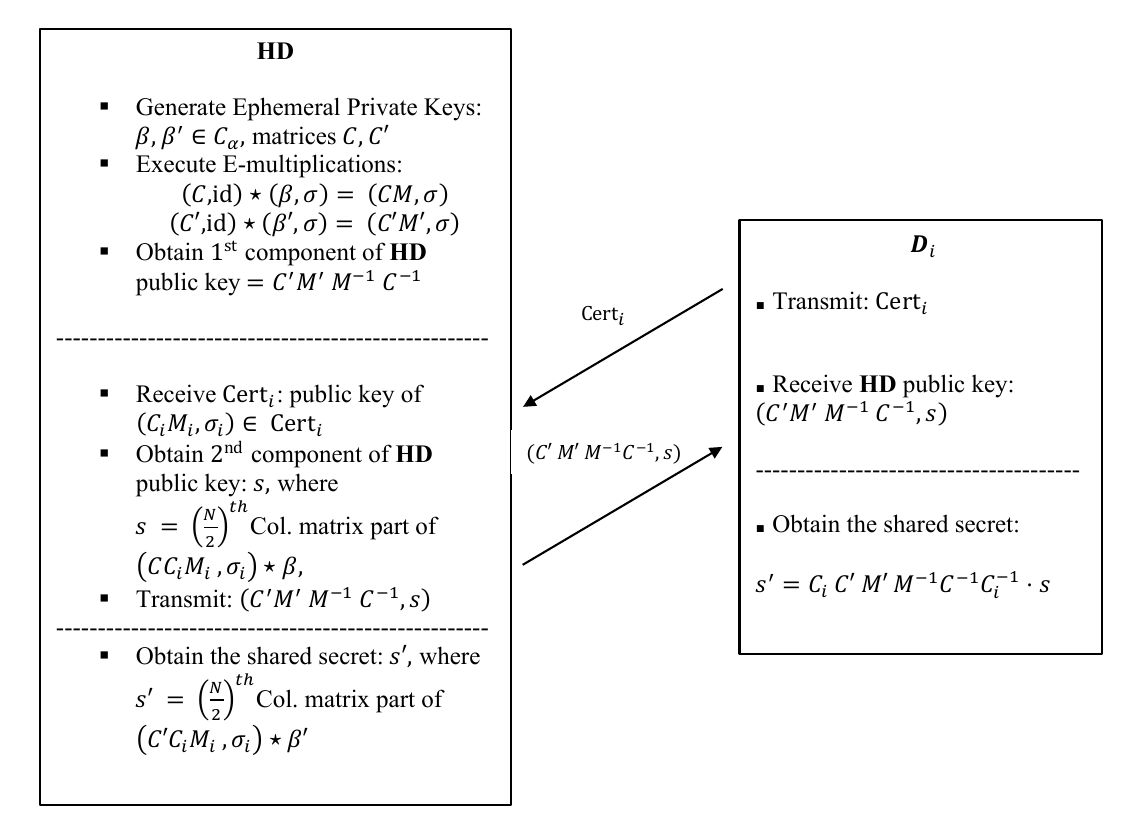}
  \caption{Ironwood Protocol Flow}
  \label{f:protocol_flows}
\end{figure}

 \vskip 10pt 
{\bf Step 1:} The device $D_i$ sends HD the certificate
$\text{Cert}_i$, which contains a copy of $\text{Pub}_i$ that has
been digitally signed by the TTP. Here $\text{Pub}_i$ is the public
key of $D_i$ and the matrix $C_i$ is the private key of $D_i$.
 
 \vskip 10pt
{\bf Step 2:} The HD generates two ephemeral non-singular matrices 
\[
C = \sum_{k=0}^{N-1} c_{k} m_0^k, \qquad C' = \sum_{k=0}^{N-1} c'_{k} m_0^k.
\]
Here $c_{k}, c'_{k}\in \F_q$ are chosen just as the coefficients in
the creation of the $C_{i}$ were.
 
\vskip 10pt 
{\bf Step 3:} The HD generates an ephemeral permutation
$\sigma$ and two ephemeral braids $\beta, \beta'$; the latter are
random words in $\mathcal C_\alpha$ as before, and we require that
$\beta$ and $\beta'$ have the same permutation $\sigma =
\sigma_{\beta} = \sigma_{\beta'}$.  This can be accomplished
efficiently by first generating a braid using the first half of
conjugates, and then create the second braid by using the same set of
conjugates and adding choices from the set of conjugates where
$\alpha_i$ are purebraids.  Alternatively, if none of the $\alpha_{i}$
are pure, one can simply take a word in the HD's conjugates to a
sufficiently high power to make it pure.

 \vskip 10pt\noindent {\bf Remark:} {\it After Step 3 the construction
of the ephemeral part of the private key of the {\rm HD}, which
consists of $C,C',\beta, \beta', \sigma$, is complete. The $T$-values
and the set of conjugates $\mathcal C_\alpha$ are also part of the
private key of the {\rm HD} and must be treated as confidential
information.}
 
  \vskip 10pt {\bf Step 4:} Using the $T$-values, the HD computes the
following two E-multiplications: \vskip -5pt 
\begin{align*}
(C, \text{Id})\star (\beta, \sigma) &:= (C M, \sigma),\\
(C', \text{Id})\star ( \beta',\sigma) &:= (C' M', \sigma).
\end{align*}
 
   \vskip 10pt
 {\bf Step 5:} The HD  has received $\text{Pub}_i = (C_i M_i, \sigma_i) $ in the signed digital signature sent by $D_i$. 
 Next, using the $T$-values, the HD computes the following two
 E-multiplications:
\begin{align*}
(C C_i M_i, \sigma_i) \star (\beta, \sigma) &:= (Y, \,\sigma_i \sigma),\\
(C' C_i M_i, \sigma_i) \star (\beta', \sigma) &:=  (Y'\hskip-1pt,  \,\sigma_i \sigma).
\end{align*}
 
 \vskip 10pt
 {\bf Step 6:} The HD computes
\begin{align*}
s &= \left(N/2\right)^{\text{th}} \; \text{column of the matrix} \; Y,\\
s' &= \left(N/2\right)^{\text{th}} \; \text{column of the matrix} \; Y'.
\end{align*}

 \vskip 10pt
 {\bf Step 7:} The HD sends $D_i$ the pair
\[
\big(C' M' M^{-1} C^{-1}, \;\, s\big).
\]
 
 \vskip 10pt {\bf Step 8:} The device $D_i$ receives the matrix $C' M'
M^{-1} C^{-1}$ and the vector $s$ and verifies that at least one half
of the entries of $s$ and the matrix are not zero (which can be done
in $N$ steps and $N^2$ steps, respectively).  These checks are to
prevent a class of invalid public-key attacks. If at least 1/2 of the
entries of either $s$ or the matrix are zero then the protocol
halts. Otherwise, the device $D_i$ computes the matrix and vector
multiplications:
\[
s' = C_i \left(C' M' M^{-1} C^{-1}\right) C_i^{-1}\cdot s.
\]
The device $D_{i}$ can do this since it knows its private key $C_i$
and has received $C' M' M^{-1} C^{-1}$ and $s$ from the HD. Further,
if $s=s'$ the protocol should also halt.
 
 \vskip 10pt
 {\bf Shared Secret:} The shared secret is the column vector $s'$,
 which is now known to both HD and $D_i.$
 
 \vskip 10pt {\bf Step 9:} The final step is to authenticate the
device $D_i$. Mutual authentication can be established by checking
that the HD and $D_i$ have obtained the same shared secret.  This is
because the device $D_i$ has sent the HD the signed certificate
containing a copy of its public key and the unique HD is the only
entity with access to the secret conjugate material and $T$-values
enabling it to produce the correct response.  Methods for doing this,
such as using a hash to create a validation value or using a nonce and
Message Authentication Code (MAC) in a challenge/response protocol,
are well known, so we do not reproduce them here.

\vskip5pt
Figure \ref{f:protocol_flows} gives a diagram that  compactly summarizes how HD and
$D_i$ arrive at the shared secret $s=s'$.

 \vskip5pt At this point the devices can mutually authenticate. The HD
can authenticate the device $D_i$ by verifying its certificate and
then having $D_i$ prove knowledge of the private matrix associated
with the public matrix in the certificate.  The device $D_i$ proves
this knowledge by showing that it can generate the same shared secret
as the HD.

 \vskip5pt In the other direction, the $D_i$ device can authenticate
the HD if the HD proves that it has created the same shared secret and
$D_i$ verifies that $C' M' M^{-1} C^{-1}\not= C_i \left(C' M' M^{-1}
C^{-1}\right) C_i^{-1}$ which is equivalent to checking that $s \ne
s'$ in step 8. The latter verification thwarts a trivial spoof of the
HD (where $\beta,\beta' = 1$).  The former condition is sufficient
because only one HD contains the conjugate data to generate the HD
keypair, so only that HD could generate a public key that would create
the same shared secret.  Further, since the protocol halts in step 8
if the vector $s$ has more than $N/2$ entries which are zero, it is
not possible for an attacker to act as the HD with an invalid vector
$s$ such as $s = 0.$.  See Section \ref{s:invalid_pubkey_attack} for
further analysis.

 \vskip 5pt 
It is not at all obvious that the column vectors $s,s'$ produced by the
HD and $D_i$ have to be equal. We now provide a proof of this.  To
begin, the braids $\beta$ and $\beta'$ commute with $\beta_i$, since
they are formed from the sets of conjugates $\mathcal C_\alpha,
\mathcal C_\gamma$, respectively, and these sets of conjugates
commute. It follows from Step 5 that 
\begin{align*}(C C_i M_i,
\sigma_i) \star (\beta, \sigma) & =
     \big(C_i CM, \sigma   \big) \star (\beta_i, \sigma_i) \\ & = (Y, \, \sigma_i \sigma), \\
     (C' C_i M_i, \sigma_i) \star (\beta', \sigma) & = \big(C_i C' M',
\sigma \big) \star (\beta_i, \sigma_i) \\ & = (Y'\hskip-1pt,
\,\sigma_i \sigma).  
\end{align*} 
Now define an unknown matrix $X$ by
the formula 
\[
(1, \sigma)\star \left(\beta_i, \sigma_i \right) =
\left(X, \sigma_i\right).
\]
It follows that 
\[
Y = C_i C M X, \qquad Y'
= C_i C'M' X.
\]
Next, define a column vector $x$ where 
\[
x = \left(C_i
CM\right)^{-1}\cdot s.
\] 
The column vector $x$ is just the
$\left(N/2\right)^{\text{\rm th}}$ column of the matrix
$X$. Hence 
\[
s' = C_i C' M' \cdot x = C_i C'M' M^{-1} C^{-1}
C_i^{-1}\cdot s,
\]
which shows that the computed secrets agree.
\section{Security Analysis of Ironwood}

  The Ironwood protocol is an outgrowth of the Algebraic
Eraser$^{\text{TM}}$ key agreement protocol (AEKAP) first published in
\cite{aagl} in 2006. The security of the AEKAP was based on the
difficulty of inverting E-multiplication and the hard problem of
solving the simultaneous conjugacy search problem for subgroups of the
braid group. The AEKAP had withstood numerous attacks (see
\cite{GKTTV}, \cite{G},\cite{GG}, \cite{Gu}, \cite{HS},\cite{MU}) in
the last 10 years. However, the recent successful attack of Ben-Zvi,
Blackburn, Tsaban (BBT) \cite{BBT}, for small parameter sizes,
requires an increase in key size (see \cite{aagg}) to make AEKAP
secure against the BBT attack.

The Ironwood protocol was designed to be totally immune to the BBT
attack \cite{BBT} without compromising on key size, speed or power
consumption. A necessary requirement for the security of Ironwood is
that the $T$-values and conjugates which are distributed to the HD
cannot be obtained by an adversary. The $T$-values and conjugates are
not on any of the other devices $D_i$ in the network. Without knowing
the $T$-values and conjugates the BBT attack \cite{BBT} cannot proceed
at all.
 
 It is also clear that the Ironwood protocol satisfies the last
requirement of an MKAAP.  Namely, if an attacker can break into one of
the devices $D_i$ and obtain its private key, then only the security
of $D_{i}$ is breached; all other devices remain
secure. This is because the only secret information on the device
$D_i$ is the private key $C_i$. Knowledge of $C_i$ has no effect on
the key agreement and authentication protocol between the HD and other
devices $D_j$ with $j \ne i.$

We now present a preliminary informal security analysis of Ironwood.

\subsection*{Reversing E-multiplication is Algorithmically Hard}

Strong support for the hardness of reversing E-Multiplication can be
found in\cite{Mulland} which studies the security of \emph{Z\'emor's
hash function} \cite{Zemor}.  This is a hash function $H\colon
\{0,1\}^* \rightarrow SL(2, \F_q)$ constructed by fixing two
matrices $h_{0}, h_{1} \in SL (2,\F_{q})$.  Then if $B$ is the
bitstring $b_{1}b_{2}\dotsb b_{n}$, one puts
\[
H (B) = \prod_{i=1}^{n} h_{b_{i}}.
\]
For example the bitstring $01101$ hashes to the product
$h_{0}h_{1}h_{1}h_{0}h_{1} \in SL (2, \F_{q})$. Zemor's hash
function has not been broken since its inception in 1991.  In
\cite{Mulland} it is shown that feasible cryptanalysis for bit strings
of length 256 can only be applied for very special instances of $h$.
Now E-Multiplication, though much more complex, is structurally
similar to a Z\'emor-type scheme involving a large finite number of
fixed matrices in $SL(2,\F_q)$ instead of just two matrices
$h_{0}, h_{1}$. Further, E-multiplication is highly non-linear (in
contrast to ordinary matrix multiplication) since it involves
permutation of variables of Laurent polynomials.  This serves as an
additional basis for the assertion that E-Multiplication is difficult
to reverse.

 \vskip 10pt
\subsection*{Invalid Public-Key Attack}\label{s:invalid_pubkey_attack}

We now consider an invalid public-key attack of the type presented in
\cite{BR}. Such an attack assumes that an adversary can impersonate
the HD and run the Ironwood authentication protocol (using invalid
public keys) with a device $D_i.$ This type of attack is thwarted in
Step 8 of the protocol if a rogue HD (i) sends an invalid vector $s$
to $D_i$, (ii) chooses $\beta, \beta'=1 $ or very short, or (iii)
sends a matrix that is mostly 0.  In all cases $D_{i}$ can look at the
matrix and ensure there are a sufficient number of non-zero entries,
in other words that the matrix is ``far'' from the identity.

We note that this attack does reduce the minimum possible security
level, because an attacker needs to only search through $q^X$ possible
states (where $X$ is the number of non-zero entries in the $s$
vector).  By requiring at least half the entries to be non-zero, we
reduce the minimum possible security level by half. If we require more
of $s$ to be non-zero we can minimize the impact at the expense of
possibly considering a real transaction to be bogus (because each
entry in $s$ has a 1 in $q$ chance of randomly being 0).  Moreover, by
requiring the HD matrix to be more than half non-zero, it forces the
result of a valid $D_i$ computation to mix the results sufficiently
such that a reduced-space $s$ vector will still incur the $q^X$ search
space, even if an attacker chooses short $\beta, \beta'$ and a
non-zero matrix.

Further, if the $D_i$ uses a hash to create a validation value that
does not reveal the shared secret in any way or the $D_i$ uses a nonce
and Message Authentication Code (MAC) in a challenge/response protocol
(see \cite{AG}) then an invalid key attack would not directly reveal
any information to a rogue HD.
  
 Consider now the reverse case where a rogue device $D_i$ is trying to
attack the HD by sending an invalid public key to the HD. If the HD
reveals $s$ to a rogue $D_i$ using an invalid public-key attack of
$D_i$ it may lead to potential leakage.  The best approach to protect
against an invalid public-key attack against the HD is to have the
device $D_i$'s public key signed by a trusted CA/TTP.  This allows the
HD to check that the public key of the device $D_i$ is valid by
validating the certificate. If the certificate is not valid the
protocol terminates. Recall that this is the method we propose in Step
1 of the Ironwood Protocol.
     
In both cases the use of single-use ephemeral keys prevent an
attack.  If an attacker works against an HD (or a $D_i$) which uses a
single-use ephemeral key, then multiple invalid-key attacks would
always return unique responses.

\subsection*{Length Attacks and Simultaneous Conjugacy Search Attacks}
  
Although AEKAP has withstood length attacks and simultaneous conjugacy
search attacks (see \cite{Gu}) of the type presented in \cite{GKTTV},
\cite{G}, \cite{HS}, \cite{MU}, these attacks completely fail for
Ironwood. This is because it is assumed that the two sets of
conjugates, $\mathcal C_\alpha, \mathcal C_\gamma$, are not known to
an adversary. These two sets of conjugates are not in memory on any of
the devices $D_i$, and only one of the sets $\mathcal C_\alpha$ is in
memory on the HD. An assumption of Ironwood is that an adversary
cannot obtain secret information stored on the HD.
    
\subsection*{A Class of Weak Keys}

     It is  crucial that $C_i$ does not commute with $M'M^{-1}$. Otherwise an adversary can compute
     $$s' = (C'M') \cdot (C M)^{-1}\cdot s.$$ Similarly, it is also crucial that $M_i$ does not commute with $(C' M')\cdot (CM)^{-1}.$ Otherwise an attacker can compute
      $$s' = (C_i M_i) \cdot (C' M')^{-1}\cdot (CM) \cdot (C_i M_i)^{-1}\cdot s.$$
   The probability that one of the above commuting occurs is very small.  An upper bound for the probability that two matrices commute in $GL(N, \F_q)$ can be determined as follows. It is well known that there are 
$$\prod_{k=0}^{N-1} (q^N - q^k)$$
elements in $GL(N, \F_q)$, denoted $\#GL(N,\F_q)$. On taking logarithms, summing over $k$, and exponentiating back, it may be shown that
$$\#GL(N, \F_q) \ge q^{N(N-1) - \frac{N}{q\log q}}$$ for $N,q \ge 8.$
For two matrices $X, Y  \in GL(N, \F_q)$ to commute, $X$ must be in the centralizer of $Y$, and for a generic matrix $X$, its centralizer consists of polynomials in $X$. The number of such polynomials is at most $q^N.$ So an upper bound for the probability that two matrices in $GL(N, \F_q)$ commute is given by
$$\frac{q^N}{q^{N(N-1) - \frac{N}{q\log q}}}.$$ For example, when $N = 16$ and $q = 256$, the upper bound for the probability is $3.815 \times 10^{-540}$.

\subsection*{Quantum Resistance of Ironwood}

The Ironwood MKAAP and underlying E-Multiplication appear resistant to
known quantum attacks.  The following sections provide an overview and
analysis.

\vskip 5pt 
\subsection*{Resistance to Shor's Quantum Algorithm}

\noindent Shor's quantum algorithm\cite{shor} enables a sufficiently
large quantum computer to factor numbers or compute discrete logs in
polynomial time, effectively breaking RSA, ECC, and DH. It relies on
the existence of a fast quantum algorithm to solve the Hidden Subgroup
Problem (HSP) when the hidden subgroup is a finite cyclic group. It is
known that HSP can be solved on a quantum computer when the hidden
subgroup is abelian\cite{hsp}.

Ironwood, but more specifically E-Multiplication, are constructions
based on the infinite non-abelian braid group. In fact, the braid
group is torsion free and, hence, has no finite subgroups.  As a
result, there seems to be no way to apply Shor's algorithm to attack
Ironwood.

\vskip 5pt
\subsection*{Resistance to Grover's Quantum Search Algorithm}

\noindent Grover's quantum search algorithm\cite{grover} allows a
Quantum computer to search for a particular element in an unordered
$n$-element set in a constant times $\sqrt{n}$ steps as opposed to a
constant times $n$ steps required on a classical computer.  Resistance
to Grover's search algorithm requires increasing the search
space. Since E-Multiplication scales linearly, this means that if an
attacker has access to a quantum computer running Grover's algorithm,
it is only necessary to double the running time of Ironwood to
maintain the same security level that currently exists for attacks by
classical computers. In comparison, the running time of ECC would have
to increase by a factor of 4 since ECC is a based on a quadratic
algorithm.
   
\subsection*{Brute Force Attacks on the Ironwood Key Agreement Protocol}

    We now discuss the security level of the individual secret components in the Ironwood protocol. For accuracy we give the following definition of {\it security level}.

\noindent
\begin{defn}
    {\bf (Security Level):} {\it A secret is said to have  security level $2^k$ over a  finite field $F$ if the best-known attack  for obtaining the secret involves running an algorithm that  requires at least $2^k$ elementary operations (addition, subtraction, multiplication, division) in the finite field $F$.}
\end{defn}

We assume that Ironwood is running on the braid group $B_N$ over the finite field $\F_q$. Note that there are $q^N$ polynomials of degree $N-1$ over $\F_q$. So a brute force search for a particular polynomial of degree $N-1$ over $\F_q$ has security level $q^N$.
\vskip 10pt
 $\bullet$ The brute force security level of the matrix $C_i$ is $q^N.$
\vskip 4pt 
 $\bullet$ The brute force security levels of  the matrices  $C, C'$ are $q^N$.
 \vskip 10pt \noindent
  The $T$-values is a set of field elements $\{\tau_1, \tau_2, \ldots, \tau_N\}$ where none of the $\tau_i$ = 0 or 1.

  \vskip 10pt
 $\bullet$ The brute force security level of the $T$-values is $(q-2)^N.$
 \vskip 10pt
 Note that the size of the public keys $\text{Pub}_i$ of the devices 
 $D_i$ is $N^2\cdot \log_2(q) + N\log_2(N)$ and the size of the public key of the HD is $(N^2+N)\cdot\log_2(q).$ We can thus assert
 \vskip 10pt
 $\bullet$ The brute force security level of the exchanged key is $2^{N\log_2(q)} = q^N.$ 
   \vskip 10pt
 $\bullet$ The brute force security level of either of the private braids $\beta, \beta'$ is  $$SL \; > \; (2r)^L$$
 where $L$ is the length of the braid as a word in the conjugates
 assigned to the HD, and hence
 we have the lower bound
 $$SL\;>\; {\text{min}}( (2r)^L, (q-2)^N).$$
\vskip5pt
An active attacker who attempts to run a weak-key attack can force a reduction in security level.  Specifically, we would expect the search-space of $s'$ to be $q^X$ where $X$ is the number of non-zero entres in the $s$ vector sent by the HD to $D_i$, or more accurately, the number of non-zero entries required by $D_i$.  An attacker who sends an $s$ vector with just under half of the entries 0 would reduce the security level by half.  Therefore, to properly defeat this kind of attack requires choices for $N$ and $q$ such that $q^N \ge 2^{2SL}$, or more accurately, $q^X \ge 2^{SL}$ where $X < N$. 
 \vskip5pt
A passive eavesdropper only gains access to the public keys and $s$-column.  That does not provide enough information to reproduce the shared secret.  In that E-Multiplication is conjectured to be a one-way function, knowledge of $(C_iM_i, \sigma_i)$ does not enable an attacker to learn $C_i$, which would be required to compute the shared secret as $D_i$.  Similarly, knowledge of $(C'M'M^{-1}C^{-1})$ does not provide enough information to deduce $C, C', \beta,$ or $\beta'$.  This prevents computing the shared secret as the HD using $D_i$'s public key.

 \vskip5pt
If an attacker breaks into one $D_i$ device and reads out its key material, they cannot use that against another device $D_j$ ($i \ne j$).  Each device's matrices are independently generated, so knowledge of one provides no information about any other device keys.
 \vskip5pt
   It has become standard in the art to give security proofs for both asymmetric key exchange protocols and digital signatures. The structure of such proofs do not lend themselves  to the Ironwood protocol (or any other MKAAP), and as of this writing no other security proof  which would has been introduced to the field.
\section{Implementation Experience}\label{s:implementation}

For testing purposes Ironwood was implemented on multiple platforms.  Because the Other Devices only need to perform a single matrix multiplication and a single vector multiplication, we focused our effort on the requirements of the HD, as those operations are more consuming and therefore more interesting to explore.

Operationally the HD needs to perform two sets of E-Multiplication operations (one with $\beta$ and another with $\beta'$), which take the majority of the execution time.  A single E-Multiplication operation in $B_N$ requires $N$ multiplies and $2N$ additions over the finite field $\F_q$.  These operations, in turn, gets multiplied by the number of Artin generators in each braid.

As an example, we generated key material using $B_{16} \F_{256}$ for a proposed $2^{64}$ security level.  We generated 32 conjugates for each set and from there generated key material for testing.  For this testing we generated 10 sets of HD keys which averaged a braid length of 2659.2 Artin Generators for $\beta$ and 4302.4 for $\beta'$.

The first platform tested was a Texas Instruments (TI) MSP430 16-bit (model) microcontroller.  This platform runs at various speeds from 8Mhz to 30Mhz (or faster).  On this platform we used the IAR (2011) compiler, version 5.40.1 with Optimizations set to High and all transformations and unrolling options checked.  With this setting the Ironwood HD implementation built into 3126 bytes of ROM and ran with 354 bytes of RAM.  Running over the 10 keys, the MSP430 required anywhere from 4,532,480 to 6,002,668 cycles with an average of 5,309,182.  At 25MHz this equates to an average runtime of 212ms.
%Compare this to Curve25519 in \cite{25519} which takes 9,119,840 cycles in 13,112 bytes of ROM and 384 bytes of RAM.
Ironwood does not require a hardware multiplier.

The second platform was an NXP LPC1768 running at 48MHz, which contains an embedded ARM Cortex M3.  We compiled our code using GCC (arm-none-eabi-gcc) version 4.9.3 using optimization level -O3.  This built down into 2578 bytes of ROM and the runtime required 1192 bytes of RAM.  Running the Ironwood shared secret calculation over the 10 keys, this ARM platform required anywhere from 1,538,472 to 2,026,216 cycles to compute a shared secret, resulting in a runtime of 32.1 to 42.2ms (averaging 37.4ms).
%Compare this to Curve25519 in \cite{25519} which required 3,589,850 cycles (on a Cortex-M0) in 7,900 bytes of ROM and 584 bytes of RAM.

\begin{table}[htb!]
  \vskip -10pt
  \label{t:performance}
  \caption{Performance on MSP430, LPC1768 (in Cycles)}
  \begin{center}\begin{tabular}{|c|c|c|c|} \hline 
      \multicolumn{2}{|c|}{\bf Artin Length} & {\bf MSP430} & {\bf LPC1768} \\
      {\bf $|\beta|$} & {\bf $|\beta'|$} & & \\ \hline
      2626 & 5272 & 6002668 & 2026216  \\
      2332 & 3580 & 4532480 & 1538472  \\
      2414 & 3944 & 4862464 & 1648742  \\
      3172 & 4266 & 5661952 & 1914009  \\
      2168 & 4514 & 5101824 & 1728545  \\
      3092 & 4698 & 5922048 & 2000312  \\
      2978 & 3968 & 5297664 & 1792959  \\
      2744 & 4420 & 5459456 & 1845502  \\
      2430 & 4762 & 5479424 & 1854446  \\
      2636 & 3600 & 4771840 & 1617670  \\
      \hline
      2659.2 & 4302.4 & 5309182 & 1796687 \\
      \hline
  \end{tabular}\end{center}
  \vskip -10pt
\end{table}

The third platform was a TI CC2650, an embedded ARM Cortex M3 running at 48MHz on TI-RTOS.  On this platform we used TI's arm compiler (listed as TI v5.2.0).  It was configured at optimization level 4 (Whole Program optimizations) with a size-speed tradeoff (SST) of 5 (ranging from 0 to 5, 0 being fully size optimized, 5 being fully speed optimized).  At this level the code used 3568 bytes of ROM and 1192 bytes of RAM.  With this setting Ironwood computed a shared secret in an average of 37.4ms.

We also performed tests using the size-speed tradeoff of 2, which resulted in a smaller code size of only 1954 bytes of ROM and resulted in a very minor speed penalty, reducing the average computation time to 37.6ms.  Note that on this platform we couldn't get a cycle count, only a timer, and the timer API only has a $2^{16}$ cycle resolution timer, which means the timer increments every $2^{16}/48^6 = 1.37$ms.  This implies the timer results are +/-0.7ms.  However the times are still on par with the timing on the LPC1768.

\begin{table}[htb!]
  \vskip -12pt
  \label{t:performance2}
  \caption{Performance on MSP430, LPC1768, CC2650 (in ms)}
  \begin{center}\begin{tabular}{|c|c|c|c|c|c|} \hline 
      \multicolumn{2}{|c|}{\bf Artin Length} & {\bf MSP430} & {\bf LPC1768} & \multicolumn{2}{c|}{\bf CC2650 48Mhz} \\
      {\bf $|\beta|$} & {\bf $|\beta'|$} & {\bf 25MHz} & {\bf 48MHz} & {\bf (SST 5)} & {\bf (SST 2)} \\ \hline
      2626 & 5272 & 240.1 & 42.2 & 42 & 42 \\
      2332 & 3580 & 181.3 & 32.2 & 32 & 32 \\
      2414 & 3944 & 194.5 & 34.3 & 34 & 35 \\
      3172 & 4266 & 226.5 & 39.9 & 40 & 40 \\
      2168 & 4514 & 204.1 & 36.0 & 36 & 36 \\
      3092 & 4698 & 236.9 & 41.2 & 42 & 42 \\
      2978 & 3968 & 211.9 & 37.4 & 37 & 37 \\
      2744 & 4420 & 218.4 & 38.4 & 38 & 39 \\
      2430 & 4762 & 219.2 & 38.6 & 39 & 39 \\
      2636 & 3600 & 190.9 & 33.7 & 34 & 34 \\
      \hline
      2659.2 & 4302.4 & 212.4 & 37.4 & 37.4 & 37.6 \\
      \hline
  \end{tabular}\end{center}
  %\vskip -25pt
\end{table}

We should note that implementations of the Ironwood Other Device are
approximately 50-times faster than the HD computations.

\section{Conclusion}

In this paper we have introduced a new concept called a Meta Key
Agreement and Authentication Protocol and defined an instance of this
protocol called the Ironwood MKAAP.  We show how it resists the range
of known attacks against E-Multiplication based protocols and how, in
addition, it is quantum resistant in that it resists both the Shor and
Grover algorithms.

Implementations of Ironwood have been built and tested on multiple platforms, and we have shown the performance numbers achieved on three different platforms leveraging two different architectures.  Specifically, we show that we can achieve a key agreement on an MSP430 in 212ms and 37ms on an ARM Cortex M3 acting as the HD. %,
																																								      %which
																																								      %is
																																								      %twice
																																								      %as
																																								      %fast
																																								      %as
																																								      %Curve25519
																																								      %using
																																								      %less
																																								      %code.

\end{document}